\documentclass[fleqn,aps,12pt,preprint,groupedaddress,showpacs,reprintnumbers,amsmath,amssymb,floatfix,nofootinbib]{revtex4-1}

\usepackage{epsfig}
\usepackage{graphics}
\usepackage{latexsym}
\usepackage{amsmath}
\usepackage{amssymb}
\usepackage{rotating}
\usepackage{subfigure}
\usepackage{bm}
\usepackage{color}
\usepackage{setspace}

\begin{document}

\title{Finite Nuclei in the Quark-Meson Coupling (QMC)  Model}
\author{J.~R.~Stone$^{1,2}$}
\author{P.~A.~M.~Guichon$^{3}$}
\author{P.~G.~Reinhard$^{4}$}
\author{A.~W.~Thomas$^{5}$}
\affiliation{ $^1$Department of Physics,\\
University of Oxford, Oxford OX1 3PU, United Kingdom}
\affiliation{ $^2$Department of Physics and Astronomy, \\
University of Tennessee, Knoxville, TN 37996, USA}
\affiliation{$^3$SPhN-IRFU, CEA Saclay, F91191 Gif sur Yvette, France}
\affiliation{$^4$Institut f\"ur Theoretische Physik II, \\
Universit\"at Erlangen, D-91058, Erlangen, Germany}
\affiliation{ $^5$ARC Centre of Excellence in Particle Physics at the Terascale and CSSM,\\
Department of Physics,University of Adelaide, SA 5005 Australia}

\begin{abstract}
We report the first use of the effective QMC energy density functional (EDF), derived from a quark model of hadron structure, to study a broad range of ground state properties of even-even nuclei across the periodic table in the non-relativistic Hartree-Fock+BCS framework. The novelty of the QMC model is that the nuclear medium effects are treated through modification of the internal structure of the nucleon. The density dependence is {\em microscopically} derived and the spin-orbit term arises naturally. The QMC EDF depends on a single set of four adjustable parameters having clear physical basis. When applied to diverse ground state data the QMC EDF already produces, in its present simple form, overall agreement with experiment of a quality comparable to a representative Skyrme EDF. There exist however multiple Skyrme paramater sets, frequently tailored to describe selected nuclear phenomena.  The QMC EDF set of fewer parameters, derived in this work, is not open to such variation, chosen set being applied, without adjustment, to both the properties of finite nuclei and nuclear matter.
\end{abstract}

\maketitle

Since the pioneering work of Vautherin and Brink~\cite{Vautherin:1971aw}, 
effective nuclear forces of the Skyrme type have proved to be a powerful 
phenomenological tool in the study of many aspects of nuclear structure  (for reviews see \cite{Bender:2003rmp,Stone:2007ppnp,Reinhard:2015}). The Skyrme energy density functional (EDF) for self-consistent mean-field models of Hartree-Fock (HF) type is derived from the Skyrme force using low-momentum expansion.  The functional contains all conceivable bilinear couplings of densities and currents up to second order in derivatives. This approach introduces 23 coupling constants (parameters) which are, in principle, density dependent. Taking a minimalistic approach \cite{Reinhard:2015} the number of the constants can be reduced to $\sim$10, which have to be fitted to empirical data, usually nuclear ground state properties. Due to  correlations in experimental data, variable sensitivity of individual parameters to data and correlations between the parameters themselves, no single optimal parameter set has yet been identified.  Presently many sets of the Skyrme EDF parameters exist, making it difficult to interpret and reliably predict nuclear properties. 

Given the power of the mean-field approach with the Skyrme EDF, we adopted this approach with a QMC EDF. In the QMC model, developed by Guichon and collaborators~\cite{Guichon:1987jp,Guichon:1995ue}, the nuclear system is represented as a collection of confined clusters of valence quarks. Using the MIT bag model~\cite{DeGrand:1975cf}, it can be shown that when the quarks in one nucleon interact {\em self-consistently} with the quarks in the surrounding nucleons by exchanging a $\sigma$ meson (a simple representation of the Lorentz scalar-isoscalar interaction known to dominate the intermediate range attraction between nucleons), the effective mass of a nucleon in medium is no longer linear in the scalar mean field ($\sigma$) and is expressed as M$^\ast_{\rm N}$ = M$_{\rm N}$  -  g$_{\rm \sigma N}  \sigma$ + $\frac{d}{2}$ (g$_{\rm \sigma N}\sigma$)$^{\rm 2}$.  By analogy with electromagnetic polarizabilities, the coefficient $d$, calculated in terms of the nucleon internal structure, is known as the `scalar polarizability' \cite{Guichon:1987jp}. The appearance of this term in the nucleon effective mass is sufficient to lead to nuclear saturation.

To clarify differences between the Skyrme and QMC EDF's, we write the QMC EDF adopted in this work $\langle H(\vec{r})\rangle =\rho M_{\rm N}+\frac{\tau}{2M_{\rm N}}+ \mathcal{H}_{\rm 0}+\mathcal{H}_{\rm 3}+\mathcal{H}_{\rm eff}+\mathcal{H}_{\rm fin}+\mathcal{H}_{\rm so}$ using notation and definitions from \cite{Guichon:2006er,Chabanat:1997} where
\begin{footnotesize}
\begin{align*}
\mathcal{H}_0+\mathcal{H}_3=\left(-\frac{3 G_{\rho }}{32}+\frac{3 G_{\omega }}{8}-\frac{G_{\sigma }}{2 \left(d  \rho  G_{\sigma }+1\right)}+\frac{G_{\sigma }}{8 \left(d \rho  G_{\sigma
   }+1\right){}^3}\right) \rho ^2+
\\   
  \left(\frac{5 G_{\rho }}{32}-\frac{G_{\omega }}{8}+\frac{G_{\sigma }}{8 \left(d \rho  G_{\sigma }+1\right){}^3}\right)
   \left(\rho _n-\rho _p\right)^2{},
\end{align*}
\begin{align*}
\mathcal{H}_{eff}=&\left[\left(\frac{G_{\rho }}{8 m_{\rho }^2}-\frac{G_{\sigma }}{2 m_{\sigma }^2}+\frac{G_{\omega }}{2 m_{\omega }^2}+\frac{G_{\sigma }}{4 M_N^2}\right) \rho
   _n+\left(\frac{G_{\rho }}{4 m_{\rho }^2}+\frac{G_{\sigma }}{2 M_N^2}\right) \rho _p\right] \tau _n+p\leftrightarrow n,
\end{align*}
\begin{align*}
\mathcal{H}_{fin}= &\left[\left(\frac{3 G_{\rho }}{32 m_{\rho }^2}-\frac{3 G_{\sigma}}{8 m_{\sigma }^2}+\frac{3 G_{\omega }}{8 m_{\omega }^2}-\frac{G_{\sigma }}{8 M_N^2}\right)
   \rho _n+\left(-\frac{3 G_{\rho }}{16 m_{\rho }^2}-\frac{G_{\sigma }}{2 m_{\sigma }^2}+\frac{G_{\omega }}{2 m_{\omega }^2}-\frac{G_{\sigma }}{4 M_N^2}\right)
   \rho _p\right]\nabla ^2(\rho _n)\\
&+p\leftrightarrow n,
\end{align*}
\begin{align*}
\mathcal{H}_{so}=&\overset{\rightharpoonup }{\nabla }.\overset{\rightharpoonup }{J}_n \left[\left(-\frac{3 G_{\sigma }}{8 M_N^2}-\frac{3 G_{\omega } \left(2 \mu
   _{\text{is}}-1\right)}{8 M_N^2}-\frac{3 G_{\rho } \left(2 \mu _{\text{iv}}-1\right)}{32 M_N^2}\right) \rho _n+  
   \left(\frac{G_{\omega } \left(1-2 \mu
   _{\text{is}}\right)}{4 M_N^2}-\frac{G_{\sigma }}{4 M_N^2}\right) \rho _p\right]\\
&+p\leftrightarrow n,
\end{align*}
\end{footnotesize}
\noindent
with $G_\sigma= g_{\sigma N}^2/m_\sigma^2$ , $G_\omega= g_{\omega N}^2/m_\omega^2$  and $G_\rho= g_{\rho N}^2/m_\sigma^2$ where $g_{\sigma N}$, $g_{\omega N}$ and $g_{\rho N}$ are free nucleon-meson coupling constants \cite {Guichon:2006er}. There are two basic differences between this and the Skyrme EDF. The QMC expresion contains, in addition to the standard point coupling terms  $\rho^2$, $\rho\tau$,  $\rho\Delta\rho$, density dependent term involving inverse powers of $(1 + d \rho G_\sigma)$ and the spin-orbit term $\propto\rho\nabla{J}$. Both the more complicated density dependence and the spin-orbit term arise naturally from the model \cite{Thomas:2004iw,Guichon:2004xg} and does not require additional parameters. 

This Letter presents a comparison of results obtained with  QMC EDF, in its present simple form, with those of a representative Skyrme EDF. The first step is to determine the four adjustable parameters of the model, the couplings $G_\sigma$,  $G_\omega$ and  $G_\rho$  and the mass $m_\sigma$ of the $\sigma$ meson (constrained as 650 MeV $<m_\sigma<$ 750 MeV). This was done in two stages. First, the parameters were constrained by properties of symmetric infinite nuclear matter at saturation, including their uncertainty. We required -17 MeV $<$ E$_{\rm 0}<$ -15 MeV, 0.14 fm$^{\rm -3}< \rho _{\rm 0}<  $ 0.18 fm$^{\rm -3}$ for the saturation energy and density and 28 MeV $<$ S$_{\rm 0}< $ 34 MeV, L  $>$ 20 MeV and 250 MeV $<$ K$_{\rm 0}<$ 350 MeV for the symmetry energy, its slope and the incompressibility. The remaining parameters of the model, the meson masses and the isoscalar and isovector magnetic
moments, which appear in the spin-orbit interaction~\cite{Guichon:1995ue,Tsushima:1997rd,Guichon:2008zz}, were taken at their physical values. The MIT bag radius R$_{\rm B}$ was set to 1 fm. This procedure yielded limits 10.2 fm$^{\rm 2} <$ G$_\sigma < $ 12.65 fm$^{\rm 2}$, 6.95 fm$^{\rm 2} <$ G$_\omega < $ 8.90 fm$^{\rm 2}$ and 6.20 fm$^{\rm 2} <$ G$_\rho < $ 8.80 fm$^{\rm 2}$.

Second, to narrow down the limits, the parameters were further constrained by specific ground state properties of selected nuclei. The QMC EDF has been incorporated into the HF+BCS code {\tt  skyax}~\cite{reinhard}, allowing for axially symmetric and reflection-asymmetric shapes. The data set consisted of selected binding energies, rms and diffraction charge radii, surface thickness of the charge distributions, the proton and neutron pairing gaps, and the spin-orbit splitting and energies of single-particle proton and neutron states, distributed across the nuclear chart.  The best parameter set was sought using the fitting protocol developed by Kl\"upfel {\it et al.}~\cite{Klupfel:2009}. In addition to the four parameters of the QMC EDF, two strengths (protons, neutrons) for volume pairing in the BCS framework~\cite{Stone:2007ppnp} were included in the fit. 

The overall best fit was given by a single set of QMC parameters: G$_\sigma$=11.85$\pm$0.02 fm$^{\rm 2}$, G$_\omega$=8.27$\pm$0.02 fm$^{\rm 2}$, G$_\rho$=7.68$\pm$0.03 fm$^{\rm 2}$ and m$_\sigma$=722$\pm$1 MeV. Note that the non-relativistic form of the functional used here differs from the relativistic form used in \cite{whittenbury2014} with the consequence the parameter values differ. The fitted parameters given above yield for nuclear matter properties E$_{\rm 0}$=-16.0$\pm0.2$ MeV, $\rho _{\rm 0}$=0.153$\pm0.003$ fm$^{\rm  -3}$, S$_{\rm 0}$=30 MeV (fixed) and L=23$\pm4$ MeV within the desired range, but, however, the value K$_{\rm 0}$=340$\pm$3 MeV considerably higher than K$_{\rm 0}\sim$220$-$240 MeV, frequently adopted in non-relativistic nuclear matter calculations. That value mainly originated from analysis of giant monopole resonance (GMR) data available in 1980's, using a Skyrme interaction \cite{Blaizot:1980}. Typical relativistic mean field calculations agree with data better at K$_{\rm 0}$ around 270 MeV. Recently Stone {\it et al.} \cite{Stone:2014}  analyzed all GMR data available to-date , in a way independent of the choice of nuclear interaction, showing that the limits on K$_{\rm 0}$ are 250 $<$ K$_{\rm 0} < $ 315 MeV. As an additional comment we note that relativistic version \cite{Stone:2007} of the QMC EDF, previously applied to cold uniform matter, showed that the contribution of a long-range Yukawa single pion exchange lowered the incompressibility from 340 MeV to $\sim$ 300 MeV, compatible with results in \cite{Stone:2014}. We intend to include the explicit pion exchange in the future development of the non-relativistic QMC model used in the present study.

The quality of the fit is summarized in the top part of Table~\ref{tab:1} and compared with the outcome of a fit with the SV-min Skyrme EDF~\cite{Klupfel:2009} performed using the same data set and analysis.  We find encouraging that the QMC EDF, with only four adjustable parameters, yielded a rms deviation 0.36\% for binding energies, comparable with 0.24\% for the Skyrme SV-min EDF. The more significant differences are in the surface thickness, diffraction radii and the neutron pairing gaps. All these properties are sensitive to details of the region around the Fermi surface where subtle differences may occur and will be further investigated.

\begin{figure}
\begin{small}
\includegraphics[scale=0.35]{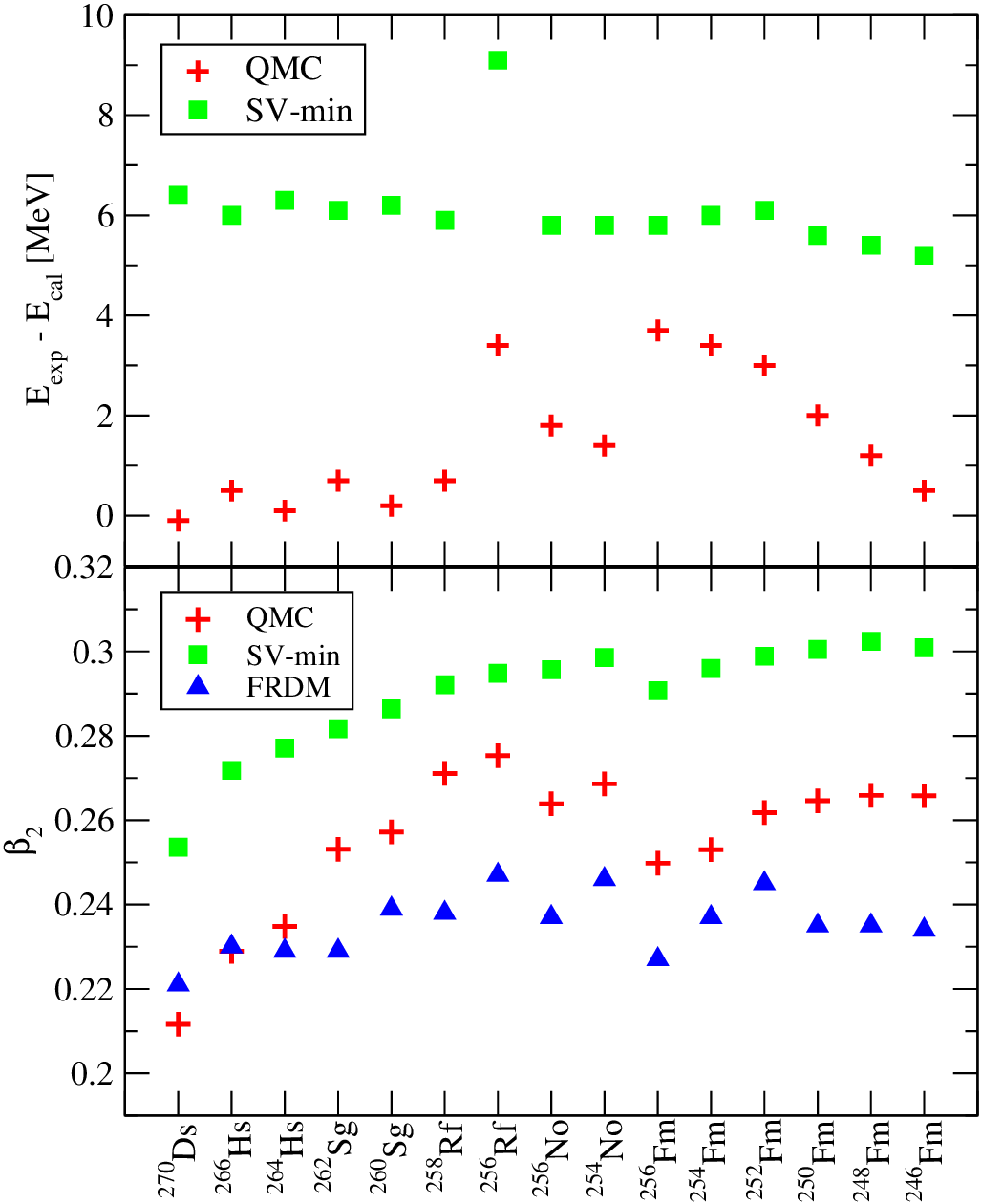}
\caption{(Color on-line) Difference between calculated and experimental ground state binding energies of SHE as obtained with QMC and SV-min EDFs (top panel). $\beta_{\rm 2}$ are shown in the botton panel which also includes FRDM \cite{Moller:1995} predictions.}
\label{fig:1}
\end{small}
\end{figure}

Using the best fit parameter set we calculated ground state binding energies of many nuclei not included in the fit. In the bottom part of Table~\ref{tab:1} rms deviations between theory and experiment for 15 super-heavy nuclei (SHE), 20 N=Z nuclei, 22 pairs of mirror nuclei and 170 spherical and deformed nuclei with known binding energy from isotopic chains with Z=38, 40, 60, 64, 86, 88, 90, 92 and 94 and isotonic chains with N=20, 28, 50, 82 and 126 are given. The most remarkable result was achieved for the SHE, where the absolute rms = 1.97 MeV for QMC and 6.17 MeV for SV-min EDF (see top panel of Fig.~\ref{fig:1}). The under-binding for SV-min EDF is a general problem in SHE for any of the standard Skyrme parametrizations \cite{Reinhard:2002cmp}. The other three groups of selected nuclei reveal that the QMC rms deviation is larger then the corresponding SV-min value by a factor less than 2. This result is encouraging considering that the QMC EDF has four parameters and SV-min EDF thirteen.

Next we examined predictions of the QMC EDF of quadrupole ($\beta_{\rm 2}$), hexadecapole ($\beta_{\rm 4}$) and octupole ($\beta_{\rm 3}$) deformation parameters. Since these parameters are not observables but are extracted from experimental data in a model dependent way, we also compare QMC and SV-min results with the Finite-Range-Droplet-Model (FRDM) of Moller {\it et al.}~\cite{Moller:1995}, which is regarded as the state-of-art benchmark for calculating nuclear masses and shapes. Where available, we use the quadrupole moment and life-time of the  I$^\pi$=2$^+_{\rm 1}$ state or a life-time related reduced transition probability B(E2, 0$^+_{\rm 1}\rightarrow$2$^+_{\rm 1}$). Indirect evidence for stable quadrupole deformation comes also a systematics of excited states (bands) built on the 0$^{\rm +}$ ground states.   

In  Fig.~\ref{fig:1} (bottom panel)  $\beta_{\rm 2}$ for SHE, as calculated in QMC, SV-min and FDRM models, are displayed. The only experimental evidence for deformation of the SHE comes from the energies of the  I$^\pi$=2$^+_{\rm 1}$ state in  $^{248-256}$Fm, $^{254}$No and $^{256}$Rf,  which all lie in the range 44$-$48 keV \cite{NNDC,Greenless:2012}, and the ratio R=E(4$^+_{\rm 1}$)/E(2$^+_{\rm 1}$) of energies of the  I$^\pi$=2$^+_{\rm 1}$ and  I$^\pi$=4$^+_{\rm 1}$  excited states. R is between 3.24$-$3.52, consistent with a stable axial rotor. The ground state bands  in $^{\rm 248,252,256}$Fm, $^{\rm 254}$No and $^{256}$Rf  show close similarity with bands observed in neighbouring U-Pu-Cm-Cf region associated with $\beta_2$=0.27$ - $0.30 \cite{Raman:2001}, in excellent agreement with $\beta_{\rm 2}$ values in Fig.~\ref{fig:1}. Thus both the ground state binding energies and the shapes of SHE predicted by QMC are in line with other models and the scant experimental evidence.

\begin{figure}
\begin{small}
\includegraphics[scale=0.35]{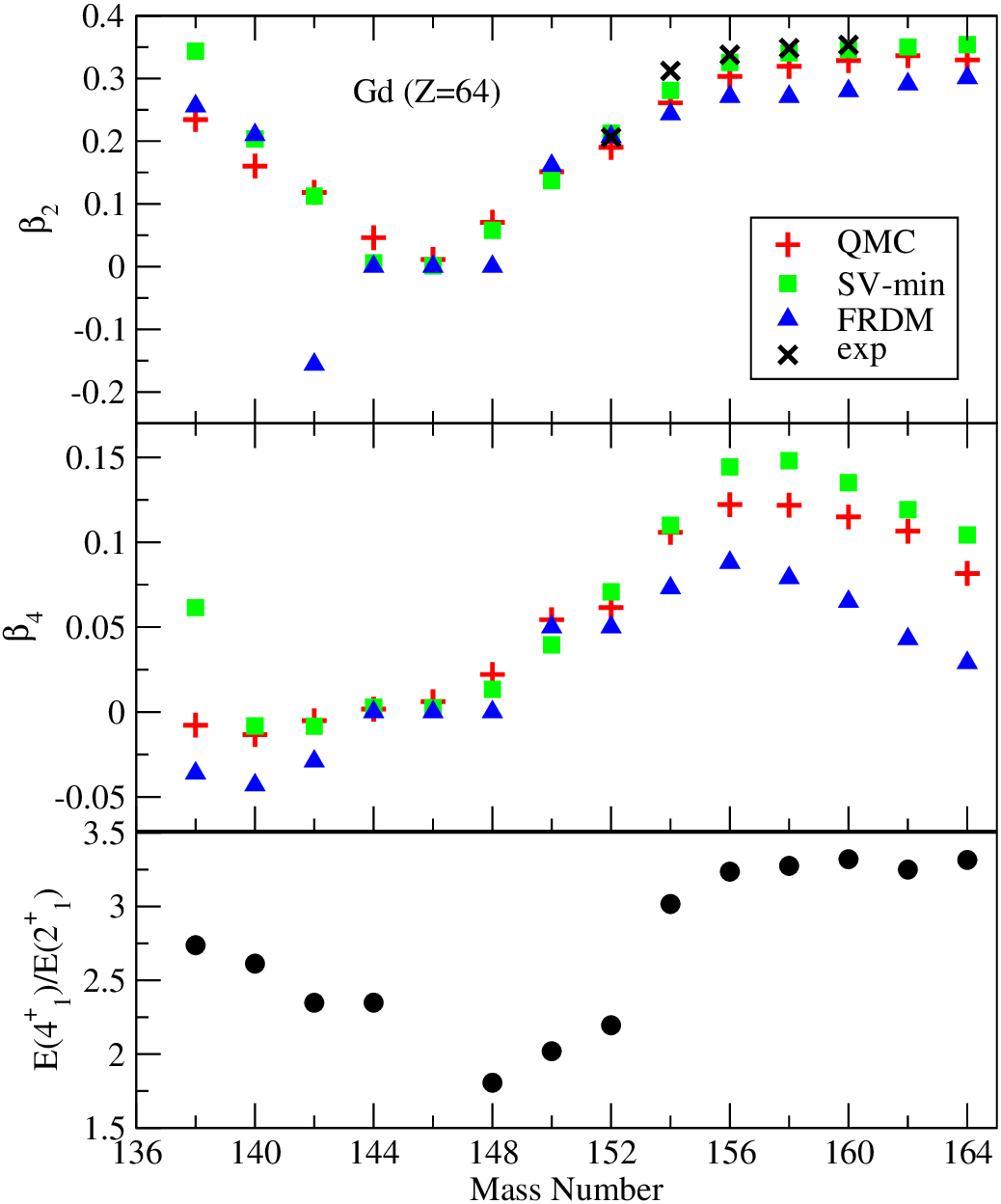}
\caption{(Color on-line) Deformation parameters $\beta_{\rm 2}$ and $\beta_{\rm 4}$ of $^{138-164}$Gd ground states (top and middle panels), obtained from QMC, SV-min and FRDM models. Experimental data \cite{Raman:2001} are added where available. Errors on $\beta_{\rm 2}$ and  $\beta_{\rm 4}$  due to uncertainties in QMC EDF parameters are less than the size of the symbol used. The ratio R = E(4$^+_{\rm 1}$)/E(2$^+_{\rm 1}$) is displayed in the bottom panel.}
\label{fig:2}
\end{small}
\end{figure}

 $\beta_{\rm 2}$ and $\beta_{\rm 4}$ calculated as a function of neutron number for the  Gd(Z=64) isotopes are presented in Fig.~\ref{fig:2}, again in comparison with SV-min and FRDM. The predictions of QMC are almost identical with the outcome of the other models. The onset and departure from collectivity is in line with the ratio R, displayed in the bottom panel. The magnitude of  $\beta_{\rm 2}$ extracted from  B(E2, 0$^+_{\rm 1}\rightarrow$2$^+_{\rm 1}$) is known in $^{152-160}$Gd \cite{Raman:2001} and the negative sign of the spectroscopic quadrupole moment Q$_{\rm s}$  of the I$^\pi$=2$^+_{\rm 1}$ state \cite{Stone:2005}  confirms the prolate shape of $^{152-160}$Gd. There is no experimental information on the value of $\beta_{\rm 4}$ but the calculation agrees well with FRDM results. 

\begin{figure}
\begin{small}
\includegraphics[scale=0.5]{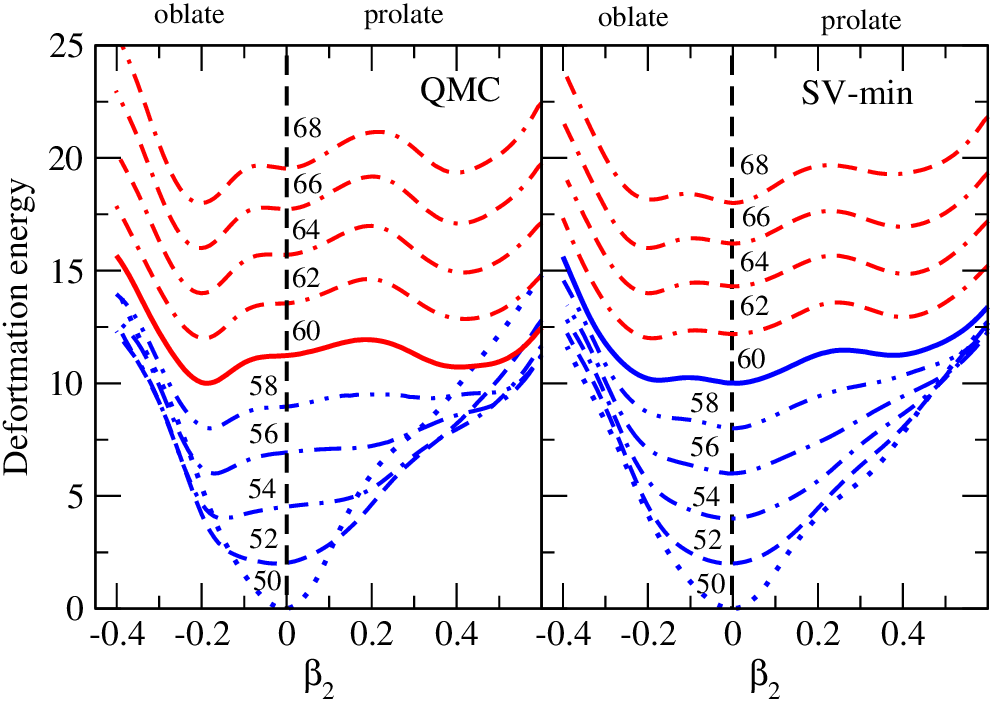}
\caption{Color on-line) Shape evolution of $^{\rm 90-106}$Zr (Z=40) isotopes as a function of neutron number obtained in constrained HF+BCS  with QMC (left panel) and SV-min (right panel) EDF. Deformation energies are displayed as a function of $\beta_{\rm 2}$. An arbitrary constant is added for each isotope, same in both panels, for display purposes. The vertical dashed line indicates the spherical shape.}
\label{fig:3}
\end{small}
\end{figure}

Fig.~\ref{fig:3} demonstrates that the QMC EDF reproduces the coexistence between spherical, oblate and prolate deformation in line with many other models of  A$\sim$100 nuclei \cite{Reinhard:1999prc,Kortelainen:2010,Rodrigues:2010,Liu:2011,Xiang:2012,Mei:2012} without additional terms or change of parameters and predicts a transition from single-particle-like structure below N=60 to collective behaviour for higher N. Very recent results results, from a Coulomb excitation experiment \cite{Sotty:2015}, show that deformation driving role in the N=60 region remains active for Z as low as 37 and that the A$\sim$100 region is still of active interest. We emphasise that the QMC EDF provides naturally the qualitative change in structure at N=60 reported in \cite{Sotty:2015}.

 An interesting suggestion, made by Dudek {\it et al.} \cite{Dudek:2014}, that a shape of a higher order tetrahedral symmetry may occur in the $^{\rm 96}$Zr ground state and compete with the quadrupole-octupole deformation, provokes the following question:  is it enough to consider the traditional prolate - oblate shapes or should one seek higher order symmetries?  The HF+BCS code used here does not have the capability to calculate them at this time. However, the suggestion of tetrahedral symmetry offers an incentive to improve the code and pursue this feature with the QMC interaction.

\begin{figure}
\begin{small}
\includegraphics[scale=0.35]{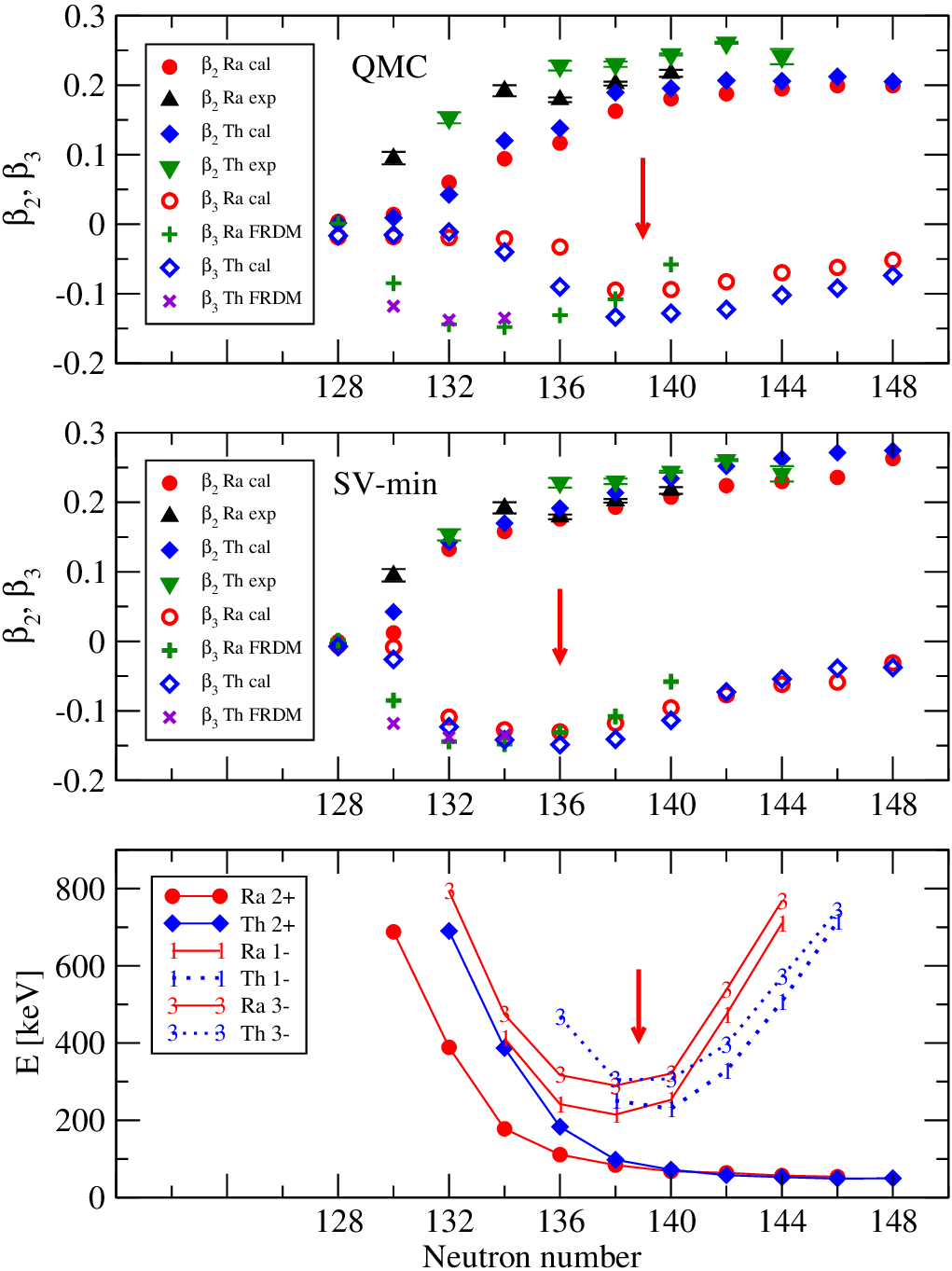}
\caption{(Color on-line) Evolution of $\beta_{\rm 2}$ and $\beta_{\rm 3}$ in Ra (Z=88) and Th(Z=90) in the range of neutron numbers 128 $\leq$ N $\leq$ 148 as predicted by the QMC (SV-min) top (middle) panel in constrained HF+BCS. Experimental data for $\beta_{\rm 2}$ are taken from \cite{Raman:2001}. FRDM values are added for comparison with $\beta_{\rm 3}$ where experimental data are not available.  The arrows indicate the maximum octupole deformation. Experimental 2$^+_{\rm 1}$, 1$^-_{\rm 1}$ and 3$^-_{\rm 1}$ excitation energies \cite{NNDC}, are displayed in the bottom panel where lines are added to guide the eye. For more explanation see text.}
\label{fig:4}
\end{small}
\end{figure}

Finally, Fig.~\ref{fig:4} demonstrates the versatility of the QMC EDF in application to quadrupole and octupole deformation in Ra and Th nuclei, again without any parameter adjustment. The evolution of $\beta_2, \beta_3$ with increasing neutron number is illustrated in the top (middle) for QMC (SV-min) EDF showing very similar trends, which also agree with that obtained for Th isotopes in ~\cite{Li:2013}. The values of $\beta_{\rm 2}$ agree with experiment \cite{Raman:2001} (where available). The scarce data on $\beta_{\rm 3}$ provide only the magnitude but not its sign \cite{Li:2013,Gaffney:2013}. However, the results in Fig.~\ref{fig:4} are supported by experimental neutron number dependence of the lowest lying  I$^\pi$=2$^+_{\rm 1}$ and  I$^\pi$=1$^-_{\rm 1}$ and  I$^\pi$=3$^-_{\rm 1}$ states in Ra and Th nuclei (bottom panel), showing close proximity of these states to the ground state for 138$\leq$N$\leq$140. The QMC model is in agreement with this data in that the maximum $|\beta_3|$ and the saturation of  $\beta_{\rm 2}$ is found at 138$\leq$N$\leq$140. SV-min and FRDM result are marginally different in predicting $|\beta_{\rm 3}|$ to reach a maximum at N=136 (see the (red) arrows in Fig.~\ref{fig:4}).

In summary, we have demonstrated for the first time that the QMC EDF, in its present form, predicts properties of even-even nuclei across the nuclear chart on a level comparable with the Skyrme EDF which has many more parameters. The novelty of this approach is that we have introduced fundamentally new physics into the EDF. Modeling the nuclear medium effect through the modification of the internal structure of the nucleon is unique to the QMC approach and has not been previously applied to nuclei to the extent reported here. That in turn led to a novel, {\it microscopically derived} density dependence. The calibrated parameters are a single, universally applicable set of four, in contrast to the larger parameter sets used in the Skyrme-Hartree-Fock mean-field models, often locally fine tuned and lacking universality. The QMC model is still at an early stage of development and the aim of this paper has been to examine its promise as compared to other models. Amongst the many levels of sophistication to consider is the inclusion of an explicit pion exchange component, known to reduce the incompressibility of nuclear matter.  It will be especially interesting to explore QMC predictions for nuclei near the limits of stability and, given its demonstrated accuracy for SHE, for potential islands of stability at very large mass number.
 
\begin{acknowledgments}
JRS and PAMG acknowledge with pleasure support and hospitality of CSSM at the University of Adelaide during visits in the course of this work. It is also a pleasure to acknowledge the technical support of R. Adorjan-Rogers during the intense computational phase of the project. This work was supported by the Australian Research Council through the ARC Centre of Excellence in Particle Physics at the Terascale (CE110001004) and by grants FL0992247 and DP150103101 (AWT).
\end{acknowledgments}

\clearpage
\begin{table}
\centering
\begin{small}
\caption{\label{tab:1} Results of the fit  yielding the parameters of the QMC EDF (top part). Experimental data selected by  Kl\"upfel et al \cite{Klupfel:2009} were used.  Equivalent results for the  Skyrme SV-min force \cite{Klupfel:2009} are added for comparison.  rms deviations of calculated ground state binding energies from experiment for four groups of nuclei, not used in the fit of parameters, are given at the bottom part of the table. They include SHE,  N=Z nuclei and N=Z$\pm$2, 4 mirror nuclei, and chains of isotopes and isotones with $|$N-Z$|$ from 2 to 60, labeled  'other'. No experimental errors were used in calculation of rms. See text for more explanation.}
\vspace{5pt}
\begin{tabular}{lcccc} 
\hline
\hline
                            & \multicolumn{4}{c}{rms deviations}\\                    
                            & \multicolumn{2}{c}{[\%]}  & \multicolumn{2}{c}{[absolute]} \\
\hline
 Data                             &     QMC      &      SV-min   &     QMC                &   SV-min        \\ \hline
Fit nuclei:                &                  &                    &                            &                       \\
Binding energies      &    0.36        &        0.24     &    2.85 MeV           &   0.62 MeV     \\
Diffraction radii        &     1.62        &         0.91     &  0.064 fm             &   0.029 fm     \\
Surface thickness     &  10.9          &        2.9       &    0.080 fm           &   0.022 fm                   \\
rms radii                 &  0.71          &        0.52      &   0.025 fm            &   0.014 fm                   \\ 
Pairing gap (n)         &  57.6          &       17.6      &     0.49 MeV          &    0.14 MeV                  \\
Pairing gap (p)         &  25.3          &       15.5       &  0.052 MeV             &  0.11 MeV\\  
Spin-orbit splitting (p)    &  15.8         &       18.5       &   0.16 MeV            &   0.18 MeV                   \\
Spin-orbit splitting (n)  &  20.3          &      16.3       &   0.30 MeV            &   0.20 MeV                   \\ 
                               &                  &                   &                             &                                   \\
Nuclei not included in the fit: &                  &                   &                             &                                   \\
Superheavy nuclei    &  0.10         &       0.32      &   1.97 MeV            &   6.17 MeV                   \\
N=Z nuclei               &       2.54           &    1.44         &     5.89 MeV          &       3.47MeV               \\     
Mirror nuclei            &      3.16           &     2.83   &    5.27 MeV           &        3.37 MeV              \\
Other                        &       0.51           &     0.30            &     4.27 MeV        &         3.19 MeV              \\  \hline
\end{tabular}
\end{small}
\end{table}
\clearpage
%
%
\clearpage
\vspace*{1cm}


\begin{thebibliography}{99}
%
\bibitem{Vautherin:1971aw} 
  D.~Vautherin and D.~M.~Brink,
  Phys.\ Rev.\ C {\bf 5}, 626 (1972).
%
\bibitem{Bender:2003rmp}
M. Bender, P.-H. Heenen, and P.--G. Reinhard,
Rev. Mod. Phys. {\bf 75}, 121 (2003).
%
\bibitem{Stone:2007ppnp}
J.R. Stone and P.--G. Reinhard,
Prog. Part. Nucl. Phys. {\bf 58}, 587 (2007).
%
\bibitem{Reinhard:2015}
P.-G. Reinhard, Phys. Scr. {\bf 91}, 023002 (2016).
%
\bibitem{Guichon:1987jp} 
P.~A.~M.~Guichon,
Phys.\ Lett.\ B {\bf 200}, 235 (1988).
%
\bibitem{Guichon:1995ue} 
P.~A.~M.~Guichon, K.~Saito, E.~N.~Rodionov and A.~W.~Thomas,
Nucl.\ Phys.\ A {\bf 601}, 349 (1996)
[nucl-th/9509034].
%
\bibitem{DeGrand:1975cf} 
  T.~A.~DeGrand, R.~L.~Jaffe, K.~Johnson and J.~E.~Kiskis,
  Phys.\ Rev.\ D {\bf 12}, 2060 (1975)
%
\bibitem{Guichon:2006er} 
P.~A.~M.~Guichon, H.~H.~Matevosyan, N.~Sandulescu and A.~W.~Thomas,
Nucl.\ Phys.\ A {\bf 772}, 1 (2006)
[nucl-th/0603044].

\bibitem{Chabanat:1997}
E. Chabanat, P. Bonche, P. Haensel, J. Meyer, R. Schaeffer, Nucl. Phys. A 627, 710 (1997)
%
\bibitem{Thomas:2004iw} 
  A.~W.~Thomas, P.~A.~M.~Guichon, D.~B.~Leinweber and R.~D.~Young,
  Prog.\ Theor.\ Phys.\ Suppl.\  {\bf 156}, 124 (2004)
  [nucl-th/0411014].

\bibitem{Guichon:2004xg} 
P.~A.~M.~Guichon and A.~W.~Thomas,
Phys.\ Rev.\ Lett.\  {\bf 93}, 132502 (2004)
[nucl-th/0402064].

\bibitem{Tsushima:1997rd} 
  K.~Tsushima, K.~Saito and A.~W.~Thomas,
  Phys.\ Lett.\ B {\bf 411}, 9 (1997)
  [Phys.\ Lett.\ B {\bf 421}, 413 (1998)]
  [nucl-th/9701047].
%
\bibitem{Guichon:2008zz} 
  P.~A.~M.~Guichon, A.~W.~Thomas and K.~Tsushima,
  Nucl.\ Phys.\ A {\bf 814}, 66 (2008)
  [arXiv:0712.1925 [nucl-th]].
%
\bibitem{reinhard}
P.~G.~Reinhard, private communication
%
\bibitem{Klupfel:2009}
P. Kl\"upfel, P.-G. Reinhard, T. J. B\"urvenich and J. A. Maruhn
Phys. Rev. C {\bf 79}, 034310 (2009)

\bibitem{whittenbury2014}
D. L. Whittenbury, J. D. Carroll,  A. W. Thomas, K. Tsushima and J. R. Stone,
Phys. Rev. C {\bf 89}, 065801 (2014)
%
\bibitem{Blaizot:1980}
J. P. Blaizot, Phys. Rep.{\bf 64}, 171 (1980).
%
\bibitem{Stone:2014}
J. R. Stone, N. J. Stone and S. A. Moszkowski, Phys. Rev. C {\bf 89}, 044316 (2014)
%
\bibitem{Stone:2007}
J. R. Stone, P.A.M. Guichon, H.H. Matevosyan, A.W. Thomas,
Nucl. Phys, A {\bf 792}, 341 (2007)
%
\bibitem{Reinhard:2002cmp}
P.--G. Reinhard, M. Bender, and J. A. Maruhn,
Comm. Mod. Phys. A {\bf 2}, 177 (2002).
%
\bibitem{Moller:1995} 
P.~Moller, J.~R.~Nix, W.~D.~Myers and W.~J.~Swiatecki,
At. Data Nucl. Data Tables {\bf 59}, 185 (1995)
%
\bibitem{NNDC}
National Nuclear Center (NNDC) hhtp://www.nndc.bnl.gov/ 
%
\bibitem{Greenless:2012}
P. T. Greenless et al., Phys.Rev.Lett. {\bf 109}, 012501 (2012)
%
\bibitem{Raman:2001}
S. Raman, C. W. Nestor, Jr., and P. Tikkanen,  At. Data Nucl. Data Tables {\bf 78}, 1 (2001)
%
\bibitem{Stone:2005}
N. J. Stone, At. Data Nucl. Data Tables {\bf 90},75 (2005)
%
\bibitem{Reinhard:1999prc}
P.-G. Reinhard, D.J. Dean, W. Nazarewicz, J. Dobaczewski, J.A. Maruhn, and M.R. Strayer,
Phys. Rev. C {\bf 60}, 014316 (1999).
%
\bibitem{Kortelainen:2010}
M. Kortelainen, T. Lesinski, J. More,  W. Nazarewicz, J. Sarich, N. Schunck, M. V. Stoitsov, and S. Wild, 
Phys. Rev, C {\bf 82}, 024313 (2010)
%
\bibitem{Rodrigues:2010}
R. Rodriguez-Guzman, P. Sarriguren, L.M. Robledo, S. Perez-Martin
Phys.Letts.B 691, {\bf 202} (2010)
%
\bibitem{Liu:2011}
Y.-X. Liu, Y. Sun, X.-H. Zhou, Y.-H. Zhang, S.-Y. Yu, Y.-C. Yang, H. Jin
Nucl. Phys. A {\bf 858} 11, (2011)
%
\bibitem{Xiang:2012}
J. Xiang, Z.P. Li, Z.X. Li, J.M. Yaoa, J. Meng
 Nucl. Phys. A {\bf 873} 1, (2012)
%
\bibitem{Mei:2012}
H. Mei,  J. Xiang,  J. M. Yao,  Z. P. Li and J. Meng, Phys. Rev. C {\bf 85}, 034321 (2012)
%
\bibitem{Sotty:2015}
C. Sotty et al, Phys. Rev. Lett.  {\bf 115}, 172501 (2015)
%
\bibitem{Dudek:2014}
J Dudek, D Curien, D Rouvel, K Mazurek, Y R Shimizu, S Tagami
Phys. Scr. {\bf 89}, 054007 (2014)
%
\bibitem{Li:2013}
Z.P. Li, B.Y. Song, J.M. Yao, D. Vretenar, J. Meng, Phys. Letts. B {\bf 726}, 866 (2013)
%
\bibitem{Gaffney:2013}
L. P. Gaffney et al., Nature {\bf 497}, 199 (2013)
%
\end{thebibliography}
\end{document}